\documentclass[12pt]{article}

\catcode`@=11
\def\secteqno{\@addtoreset{equation}{section}%
\def\theequation{\thesection.\arabic{equation}}}
\catcode`@=12
\secteqno
\usepackage{amssymb,amsfonts}

\newcommand{\be}{\begin{equation}}
\newcommand{\ee}{\end{equation}}
\newcommand{\bea}{\begin{eqnarray}}
\newcommand{\eea}{\end{eqnarray}}
\newcommand{\bref}[1]{(\ref{#1})}
\newcommand{\nn}{\nonumber}
\newcommand{\A}{\alpha} \newcommand{\B}{\beta} 
 \newcommand{\D}{\delta} 
\newcommand{\ep}{\epsilon} \newcommand{\vep}{\varepsilon}
\newcommand{\T}{\theta}

           \newcommand{\s}{\sigma}
          \newcommand{\w}{\omega}
          
\newcommand{\h}{\eta}

\newcommand{\DELTA}{\tilde\Gamma}

\newcommand{\Tb}{{\overline\theta}}

\def\6{\partial}
\def\7{\tilde}\def\t{\widetilde}
\def\8{\hat}
  \def\bD{{\bf D}}
\def\vs{\vskip 3mm}

\def\={{\quad=\quad}}\def\+{{\quad+\quad}}\def\-{{\quad-\quad}}

\def\pa{\partial}

\def\CC{{\cal C}}\def\CL{{\cal L}}

\def\bP{{\rm{\bf P}}}
\def\bL{{\rm{\bf L}}}
\def\bef{{\rm{\bf e}}}
\def\bv{{\rm{\bf v}}}\def\bw{{\rm{\bf w}}}
\def\l{{\ell}}

\def\AdS5{{AdS$_5$}}\def\S5{{ S$^5$ }}
\def\su224{{ SU(2,2$|$4)/(SO(4,2)$\times$SO(6)) }}

\def\bP{{\rm{\bf P}}}

\newcommand{\slG}{/ {\hskip-0.20cm{\Gamma}}}
\def\G11{\Gamma_{11} }
\def\NR{non-relativistic }

\def\={{\;=\;}}\def\+{{\;+\;}}\def\too{{\;\to\;}}\def\-{{\;-\;}}

\begin{document}
\thispagestyle{empty}

\hfill {\small UB-ECM-PF-05/12, KUL-TF-05/15, Toho-CP-0577}

\vskip 15mm
\begin{center}
{\Large\bf Non-Relativistic Superstrings:} 
\vskip 3mm 
{\Large\bf A New Soluble Sector of AdS$_5\times$ S$^5$}

\vskip 6mm
\vskip 4mm
\vskip 6mm

\parskip .15in

{\bf Jaume Gomis}\\
{\it  Perimeter  Institute for Theoretical Physics\\Waterloo,
Ontario N2L 2Y5, Canada}\\ Email: {jgomis@perimeterinstitute.ca }

{\bf Joaquim   Gomis}\\
{\it  Departament   ECM,   Facultat   de F{\'\i}sica,   \\
Universitat de Barcelona and Institut de F{\'\i}sica d'Altes Energies,
\\     Diagonal      647, E-08028     Barcelona,  Spain }    \\  
Email:
{gomis@ecm.ub.es }\\  and\\ 
{\it Institute for Theoretical Physics KU
Leuven\\ Celestijnenlaan 200D, B-3001 Leuven Belgium }

{\bf Kiyoshi Kamimura} \\
{\it  Department of Physics, Toho University\\ 
Funabashi\          274-8510,              Japan}\\
Email:
{kamimura@ph.sci.toho-u.ac.jp}

\medskip
\end{center}
\vskip 10mm
\begin{abstract}{
We find a new sector of string theory in AdS$_5\times$  S$^5$ describing 
non-relativistic superstrings in that geometry. The worldsheet theory of 
non-relativistic strings in  AdS$_5\times$  S$^5$  is derived and shown to reduce
 to a supersymmetric  free field theory in AdS$_2$. Non-relativistic string theory provides a new calculable setting in which to 
study holography.
 }
\end{abstract}

\parskip=7pt

\newpage%

\section{Introduction}\label{sec.1}
\indent

Recent progress in string theory suggests that  in order to define the
theory non-perturbatively,    we   must specify  the  symmetries    at
asymptotic infinity,  where observables can  be  defined. Examples of
   geometries where  holographic duals can  be constructed are
asymptotically   Anti-de  Sitter  and  asymptotically linear   dilaton
backgrounds. The holographic field theory description encodes
 the complete non-perturbative string theory dynamics in these backgrounds. 
 In practice, however, even reproducing perturbative string  physics is hampered by
 the difficulties  in solving the  worldsheet theory.

Significant progress  can be made  in solving string theory in various
backgrounds by considering a sector of the  theory that decouples from
the rest of the degrees of freedom in a suitable limit. Such decoupled
sectors are characterized by having an altogether different asymptotic
symmetry compared to that  of the parent  string theory. A  well known
example  of such a truncation    is the BMN   \cite{Berenstein:2002jq}
sector\footnote{The relevant symmetry of the BMN sector is a super-Heisenberg algebra, which   
 is   a  particular   In\"on\"u-Wigner
contraction of the $SU(2,2|4)$ symmetry  of AdS$_5\times$ S$^5$.  See
for example \cite{Hatsuda:2002xp}.} of  string theory in AdS$_5\times$
S$^5$.  Once   a  consistent sector is   found, a  complete worldsheet
theory  with the appropriate symmetries can  be written down without further
reference of the parent theory. Solving for the spectrum and interactions 
in the BMN sector has led to a concrete understanding of how perturbative string theory is encoded in the dual field theory and has shed new light on    holography in the stringy regime.

Non-relativistic string theory \cite{Gomis:2000bd} (see also
\cite{Danielsson:2000gi})   in flat space is another example of a consistent 
decoupled sector  of bosonic string  theory whose worldsheet conformal
field theory  description \cite{Gomis:2000bd} possesses the appropriate
Galilean symmetry. Non-relativistic string theory can be derived as a certain 
decoupling limit of the 
original relativistic theory, even though -- as in the BMN  case -- the 
theory can be written down without further reference to the original parent 
theory\footnote {For instance, the action of non-relativistic bosonic 
string theory can be 
obtained from  the method of non-linear realizations as a WZ term  of the
appropriate Galilean group
\cite{Brugues:2004an}.}. The basic idea behind the decoupling limit is to 
take a particular  {\it non-relativistic} limit in such a way that only states 
satisfying a Galilean invariant dispersion relation have finite energy, while 
the rest decouple. This can be accomplished
by considering wound strings in the presence of a background B-field and  tuning the B-field so that the energy coming from the background B-field cancels the tension of the string. The corresponding worldsheet conformal field theory
captures, in the usual fashion,  the
spectrum and interactions of these states. The computation of interactions among non-relativistic strings, however,  simplifies considerably compared to the relativistic case, since the path integral of the worldsheet theory ``localizes" \cite{Gomis:2000bd}  to points in the moduli space where 
holomorphic maps from the worldsheet to the target space exist.
 An analogous non-relativistic limit can be taken 
in the various corners of M-theory such that the non-relativistic states of M-theory are mapped to
each other by the action of the duality symmetries of  M-theory; thus giving rise to  a rich duality web but in 
the non-relativistic setting.

In this paper we consider non-relativistic superstring\footnote{The 
non-relativistic limit has been extended to the superbranes 
in flat space  in
\cite{Gomis:2004pw}, where the action for non-relativistic superstrings can 
be found. The symmetry of non-relativistic superstrings can be
obtained as  an   In\"on\"u-Wigner  contraction    of the ${\cal     N}=2$
super-Poincare symmetry of flat space.} theory in AdS$_5\times$ S$^5$. The 
primary motivation for studying this new limit of string theory  is to 
isolate a simple sector of the AdS/CFT correspondence
which is 
amenable to exact analysis. This should provide an interesting new 
arena in which 
to approach in a calculable setting the ideas of holography 
with ${\cal N}=4$ SYM. The main result of the paper is in fact to show that  
non-relativistic string theory in AdS$_5\times$ S$^5$
is described by a supersymmetric two-dimensional sigma model of 
free massless and massive
bosons and free massive
fermions  propagating in AdS$_2$, so that this sector is described 
by a  {\it free} theory! We show that the non-relativistic string action 
in AdS$_5\times$ S$^5$ has the supersymmetric Newton-Hooke group \cite{BacryLL} symmetry, 
which is a supersymmetrization of the kinematical group describing non-relativistic particles
in AdS. We  make preliminary comments about 
quantization of this theory and the correspondence with the dual gauge 
theory leaving a more exhaustive analysis for the future \cite{workin}.

The plan of the rest of the paper is as follows. In section two we introduce 
the basic ingredients of the Green-Schwarz action that are necessary to 
derive the worldsheet theory of non-relativistic string theory in 
AdS$_5\times$ S$^5$. Section three explains the nature of the 
non-relativistic limit of string theories   and the precise non-relativistic 
limit in AdS$_5\times$ S$^5$ is presented. 
In section four we implement the non-relativistic limit directly in the 
worldsheet action and derive the formula for the worldsheet action describing 
non-relativistic strings. In section five we fix $\kappa$-symmetry and   
show that the worldsheet action becomes {\it free}! Section six shows that 
non-relativistic string theory in AdS$_5\times$ S$^5$ is related by 
T-duality to a five-dimensional time dependent
pp-wave. Section seven contains preliminary remarks about the 
description 
of non-relativistic strings in the dual ${\cal N}=4$ theory. We have relegated 
to the various appendices useful formulae that are used in the main text.

\vs
\section{Superstring action in AdS$_5\times$ S$^5$}

In this section we introduce the basic ingredients that are needed to derive the worldsheet theory describing non-relativistic strings in AdS$_5\times$ S$^5$ as a limit of the sigma model on AdS$_5\times$ S$^5$. The worldsheet Lagrangian of  type  IIB string  theory in an  arbitrary
curved background is given  by 
\be
\CL =-{T\over 2}\left(\CL^{(Kin)}+2\CL^{(WZ)}\right),
\label{totLag}
\ee
where $T$ is the string tension, $\CL^{(Kin)}$ is the kinetic term and
$\CL^{(WZ)}$ is  the  Wess-Zumino term.  The  corresponding  action is
invariant   under worldsheet diffeomorphisms,  space-time  supersymmetry  and, when   the  type IIB
supergravity constraints are satisfied, under $\kappa$-symmetry.

The  worldsheet  action in the   AdS$_5\times$  S$^5$ background can be
formulated in  a manifestly   supersymmetric  way --   invariant  under
$SU(2,2|4)$ -- by writing it in terms of  Cartan's 
one forms on the  coset space
\be
{SU(2,2|4)\over SO(4,1)\times SO(5)}
\label{cosetrep}
\ee
describing
the AdS$_5\times$     S$^5$  superspace.  
They   are given by
\be
\Omega=-ig^{-1}dg=P_m\bL^m+P_{m'}\bL^{m'}+Q_{\alpha\alpha' I}\bL^{\alpha\alpha' I}+{1\over 2}M_{mn}\bL^{mn}+{1\over 2}M_{m'n'}\bL^{m'n'},
\label{forms}
\ee
where 
$P_{m}(P_{m'})$ denote    the   translation generators  in
AdS$_5$(S$^5$), $M_{mn}(M_{m'n'})$ are the corresponding rotation
generators and $Q_{\alpha\alpha' I}$ are the generators of supersymmetry.
 $m(\alpha)$ is a  vector(spinor) index of the $SO(4,1)$ tangent
space symmetry of AdS$_5$, $m'(\alpha')$ a vector(spinor) index of the
$SO(5)$ tangent space symmetry of  S$^5$ and $I$  is a vector index of
the $SL(2,R)$ symmetry of type IIB supergravity. In this language, the
$SU(2,2|4)$ algebra is encoded in Cartan's equation\footnote{See Appendix \ref{appNRalg} for the commutation relations.}:
\be
d\Omega+i\Omega\wedge\Omega=0.
\label{MCeq}
\ee

In order  to construct an action that  is invariant under $SU(2,2|4)$,
all that is required  is to construct out of the Cartan one forms invariants
under the stability\footnote{This follows from the well known
fact that Cartan's one forms on a coset space  $G/H$ are not invariant
under   the left action of  $G$.   The left  action  of  $G$ induces a
transformation such that  the one forms  transform
 as gauge connections of $H$, 
 that is $\Omega\rightarrow h\Omega h^{-1}-ihdh^{-1}$, where
$h\in  H$.} subgroup $SO(4,1)\times SO(5)$. One   is left  with
\cite{Metsaev:1998it}
\bea
\CL^{(Kin)}&=&\sqrt{-h}h^{ij}\left(\eta_{mn}\bL_i^m\bL_j^n+\delta_{m'n'}\bL_i^{m'}\bL_j^{n'}\right)
\nn\\
d(\CL^{(WZ)}d^2\xi)   &=&i\;\bar     \bL(\Gamma_m    \bL^m+\Gamma_{m'}
\bL^{m'})\tau_3\bL,
\label{covalg}
\eea
where   $\bL^{m,m'}_i$  is  the  pullback of  the    one form  on  the
worldsheet, $h_{ij}$ the auxiliary two dimensional metric and $\xi^i$
the worldsheet   coordinates.  The corresponding  action is  invariant
under  worldsheet   diffeomorphisms,  local Weyl transformations, $SU(2,2|4)$     and       $\kappa$-symmetry.\footnote{The coefficient of $\CL^{(WZ)}$ in \bref{covalg}\ is
determined by enforcing  that the complete action \bref{totLag} is 
invariant under $\kappa$-symmetry.}


We also find it useful to consider the worldsheet action in the 
Nambu-Goto formulation. 
 The Nambu-Goto action is  obtained
by integrating out the 
auxiliary metric $h_{ij}$ in \bref{covalg}. In doing this, only the  kinetic term is modified  and in the normalization 
used in \bref{totLag} reads
\be
\CL^{(Kin)}=2\sqrt{-\det G_{ij}},
\label{NGaction}
\ee
where  the induced metric is:
\be
G_{ij}=\h_{mn}\bL^{m}_i\bL^{n}_j+\delta_{m'n'}\bL^{m'}_i\bL^{n'}_j.
\label{imetcov}
\ee

We can turn 
one  more coupling on the  worldsheet consistent  with all
the symmetries of the Green-Schwarz  action. From the space-time point
of view, it corresponds to  turning on a  closed $B$-field, which does not modify 
the supergravity equations of motion.

 Therefore,
the worldsheet Lagrangian we are going to analyze is
\be
\CL =-{T\over 2}\left(\CL^{(Kin)}+2\CL^{(WZ)}+2\CL^B\right)
\label{totLagB}
\ee
with
\be
\CL^B=f^*B,
\label{Bfield}
\ee
where $f^*B$ is the pullback  of $B$ onto the  worldsheet.  This   coupling   is useful    in
constructing the    worldsheet action of non-relativistic   string theory  in
AdS$_5\times$ S$^5$.

In the Nambu-Goto formulation, worldsheet diffeomorphisms give rise to the following first class constraints
\bea
&& H\equiv{1\over 2T}\left[\eta^{mn} {\7p_{m}\7p_{n}}+\delta^{m'n'}{\7p_{m'}\7p_{n'}}+
T^2\left(\eta_{mn}\bL_1^m\bL_1^n+\delta_{m'n'}\bL_1^{m'}\bL_1^{n'}\right)\right]
=0, \nn\\
&&
T\equiv \7p_{m}\bL_1^m+\7p_{m'}\bL_1^{m'}=0,
\label{constWZ1}
\eea
where $\7p_\cdot\equiv{\partial {\cal L}^{(Kin)}/ \partial \bL^\cdot_0}$
 corresponds to the mechanical momentum.\footnote{
The canonical momentum differs from the mechanical momentum $\7p$ by an additive amount proportional to the $B$ field and by the contribution from the WZ term, 
generalizing the usual point particle relation $p=\7p+A$, 
where $A$ is a background  gauge field.}  We note that the constraint $H\equiv 0$ is quadratic in the energy, which is   characteristic of relativistic theories. As we show in this paper,  the corresponding constraint for non-relativistic string theory \bref{constNR} is  linear in the energy, which is   characteristic of non-relativistic theories. There are additional fermionic first class constraints associated with   $\kappa$-symmetry, whose explicit  expression we do not need in the paper.

In the Polyakov formulation, if we fix  worldsheet reparametrizations  
by choosing the   conformal
gauge      $\sqrt{-h}h_{ij}=\eta_{ij}$, the   
diffeomorphism constraints \bref{constWZ1} imply  the familiar Virasoro conditions 
\bea
G_{00}+G_{11}&&=0
\nn\\
G_{01}&&=0,
\label{const}
\eea
where $G_{ij}$ is the induced metric given in \bref{imetcov}.

Thus far we have written the action without a specific
choice of    supercoordinates  on the coset,   or equivalently we  did not
 invoke a specific choice  of coset representative $g$. In the
next section, we    choose a  parametrization which is
conducive to taking the non-relativistic limit.

\section{Non-relativistic limit of AdS$_5\times$S$^5$}\label{sec.2}

In this section we find a limit of string theory describing  non-relativistic strings in AdS$_5\times$S$^5$.
Non-relativistic string theory\footnote{In this section we briefly summarize some basic facts about   
 non-relativistic strings (see \cite{Gomis:2000bd} for more details and explanations).} can  be defined  by taking a  suitable
non-relativistic  limit of relativistic  string  theory. The basic idea is to 
find a limit  where the ``light" strings satisfy a non-relativistic dispersion relation 
while 
the rest of the degrees of freedom become infinitely ``heavy". 
This can be accomplished
by considering wound (charged) strings in the presence of a background B-field and  tuning the B-field so that the energy coming from the background B-field cancels the tension of the string.\footnote{The same limit can be taken for open strings ending on a
D-brane,  giving rise to the NCOS theory \cite{Seiberg:2000ms}\cite{Gopakumar:2000na}. It was in this context where the  appearance  of non-relativistic strings first surfaced \cite{Klebanov:2000pp}. One can also consider open strings on a D-brane transverse to the $B$-field, which gives rise to 
non-relativistic open strings \cite{Danielsson:2000mu}.}
In this way, 
only closed strings with positive charge  remain ``light", while strings with non-positive charge become infinitely ``heavy" in the limit. 
Therefore, the limit eliminates the anti-particles, whose absence is characteristic of non-relativistic theories. The winding number plays the role of the mass of the non-relativistic string.

The remaining  ``light" strings  obey  a non-relativistic dispersion relation which can be derived by either solving the Virasoro constraints 
\cite{Gomis:2000bd} or by 
solving the diffeomorphism constraints \cite{Garcia:2002fa}, which are linear in the energy and thus non-relativistic in nature. Furthermore, in this limit
the speed of light in the directions transverse to the string is  sent to infinity,  giving rise to instantaneous
action  at  a  distance,   which  is  also a  characteristic  property   of
non-relativistic point  particle  theories. 

In order to take the limit, we must specify the coordinates we are going to use to parametrize the target space. The coordinates of AdS$_5\times$S$^5$  are determined by a choice of parametrization of the $SU(2,2|4)/(SO(4,1)\times SO(5))$ coset space.  We   consider   the  following one
\be
g\=e^{iP_1           X^1}e^{iP_0X^0}e^{iP_a             X^a}e^{iP_{m'}
X^{m'}}e^{iQ_{\A\A'I}\T^{\A\A'I}},
\label{coset}
\ee
where    the  index $m$  splits  into   $(\mu,a)$   with $\mu=0,1$ and
$a=2,3,4$. Therefore, the coordinates  of AdS$_5\times$ S$^5$ are given
by  $(X^\mu,X^a)$  and $X^{m'}$ respectively,  while $\theta^{\A\A'I}$
are  the corresponding  fermionic coordinates.  This parametrization makes manifest an  embedding of AdS$_2$ in AdS$_5$, where
$X^\mu$ are the coordinates of AdS$_2$. A specific choice
 of an AdS$_2$ representative is  selected  to make manifest in the associated AdS$_2$ metric a space-like isometry  along which the non-relativistic strings can wind. 
 
Using  the  parametrization \bref{coset}  and the $SU(2,2|4)$ algebra,
we can compute the invariant one forms \bref{forms}. The
metric of AdS$_5\times$S$^5$  is obtained from 
the      bosonic  components of    the
one-forms associated with translations, which correspond geometrically to the metric vielbeins on the coset\footnote{We recall that  the   bosonic component  of the
one-forms $\bL^{mn}, \bL^{m'n'}$ associated with the generators of the
stabilizer group of the   coset, correspond geometrically to  the spin
connections    $w^{mn}(w^{m'n'})$        of  AdS$_5($S$^5$),     where
$\bL^{mn}=w^{mn}+\hbox{fermions}$                                  and
$\bL^{m'n'}=w^{m'n'}+\hbox{fermions}$.}:
\bea
\bL^m=\bef^m+\hbox{fermions}\ \Longrightarrow ds^2_{AdS}=\bef^m\bef^m
\nn\\
\bL^{m'}=\bef^{m'}+\hbox{fermions}\ \Longrightarrow ds^2_{S}=\bef^{m'}\bef^{m'}.
\eea

Using the  formulae in Appendix \ref{appLIF},  the  metric of  AdS$_5$ in these
coordinates is given by
\bea
ds^2_{AdS}&=&\cosh^2\rho     \left[    -(dX^0)^2+\cos^2\left({X^0\over
R}\right)(dX^1)^2\right]
\nn\\
&+&\left[ \left({\sinh \rho\over \rho}\right)^2(dX^a)^2-\left({\sinh^2
\rho\over \rho^2}-1\right){(X_adX^a)^2\over {\rho}^2R^2}\right],
\label{metric}
\eea
where $\rho=\sqrt{X_aX^a}/R$ and $R$ is the radius of AdS$_5$ (S$^5$),
while   the    metric inside    the   first    bracket  is    that  of
AdS$_2$ (See Appendix \ref{AppEmbed} for details of the corresponding geometry). We note that in these coordinates, ${\partial \over \partial X^1}$ is a(n almost everywhere) space-like Killing vector, so that a string can wrap along this coordinate. By appropriately tuning the background $B$-field the wound closed strings can be made ``light" and non-relativistic. 

Likewise, the S$^5$ metric in the coordinates we are using is
given by
\be
ds^2_{S}=\left({\sin   r\over     r}\right)^2(dX^{m'})^2-\left({\sin^2
r\over r^2}-1\right){(X_{m'}dX^{m'})^2\over {r}^2R^2},
\ee
where $r=\sqrt{X_{m'}X^{m'}}/R$.

The   non-relativistic   limit  rescales the   longitudinal coordinates
$X^\mu$ differently  than the  rest.  Physically, this  corresponds to
analyzing   a   string    that spans   an   AdS$_2$   subspace  inside
AdS$_5\times$S$^5$, such as a string winding along the $X^1$ coordinate.  In  order  to  get  a supersymmetric   and  $\kappa$-symmetric  action we  must  also 
appropriately rescale the fermionic
coordinates in  a way consistent with  the symmetries preserved by the
bosonic scaling. These conditions single  out a unique  scaling for $\theta^{\alpha\alpha' I}$.

In  order to obtain a well  defined Green-Schwarz action after taking the
non-relativistic limit, we introduce a closed B-field along AdS$_2$. As explained above,  
the limit involves tuning the  $B$-field so that it cancels the divergent contribution to the energy coming from the area of the worldsheet and the WZ term, so that one is  left with
non-relativistic strings of  finite energy. The
explicit     worldsheet  coupling    \bref{Bfield}  is given by
\be
{\cal L}^B=B_{\mu\nu}\bv^{\mu}_0\bv^{\nu}_1,
\label{BfieldA}
\ee
where  $\bv^\mu$  are  the  zweibeins\footnote{The  AdS$_2$  zweibein one forms 
$\bv^\mu$  can  be related to the AdS$_5$ vielbeins $\bef^\mu$ 
by   $\bef^\mu=\cosh\rho\;  \bv^\mu$.} of  the AdS$_2$ subspace
spanned by $X^\mu$ and $B_{\mu\nu}$ is a constant anti-symmetric  matrix.

Having motivated the physics, we can now write down the 
non-relativistic  limit  of string theory  in AdS$_5\times$S$^5$. It is given by:
\bea
X^\mu&=&\w x^\mu,
\qquad 
\theta=\sqrt{\w }\:\theta_-+{1\over \sqrt \w }\: \theta_+
\nn\\
B_{\mu\nu}&=&\epsilon_{\mu\nu},
\qquad \qquad 
R=\w R_0
\label{limit}
\eea
and then take $\w \rightarrow \infty$. 

In the non-relativistic limit, the longitudinal
coordinates  of the string are rescaled  while the transverse ones are
left untouched, which implies that the transverse fluctuations are small.
This also has the effect of making the speed of  light in
the   transverse   directions\footnote{The speed of light along the string is unchanged.} very  large, infinite in the limit.
 The radius $R$
is also sent to infinity,  which has the  effect of flattening out the
transverse space to  AdS$_2$. The $B$-field is tuned in such a way that  
its contribution to the energy is precisely cancelled by the contribution coming from the
tension of the string.  We shall see later that there is
a very precise sense in which the limit is non-relativistic. As we shall see, the worldsheet theory that we obtain by taking the limit \bref{limit} is invariant under the supersymmetric version of the Newton-Hooke group \cite{BacryLL}, which captures the non-relativistic kinematics of slowly moving particles in AdS. 

 The scaling behaviour  of the fermionic coordinates in \bref{limit} is characterized by
their transformation properties under  the matrix  $\Gamma_{*}$, which
splits the fermionic coordinates into two eigenspaces with eigenvalues
$\pm 1$
\be
\Gamma_*\theta_{\pm}=\pm\theta_{\pm}\qquad \hbox{where}\qquad  \Gamma_*=\Gamma_0\Gamma_1\tau_3
\ee
and $\Gamma_*^2=1$.  As shown in Appendix \ref{appNGA},  the matrix  $\Gamma_{*}$  is the first  term in the
non-relativistic expansion 
of  the matrix $\Gamma_\kappa$ appearing in the
$\kappa$-symmetry transformations of the relativistic string action. 

Having physically motivated the limit, we are now ready to analyze its consequences. 

\section{Non-relativistic string theory}\label{sec.2.1}

In this section we  derive  the worldsheet action of   non-relativistic
strings in  AdS$_5\times$S$^5$  by taking  the
 non-relativistic limit  of   the  Green-Schwarz action  in
AdS$_5\times$S$^5$ described  in  the previous section. The rigid and local symmetries of the
resulting action are also discussed and the role played by the supersymmetric extension of the Newton-Hooke group -- the kinematical group of non-relativistic AdS -- is explained. 

The basic idea is to take the non-relativistic limit \bref{limit} directly 
in the worldsheet action and obtain in this way a definition of the non-relativistic theory. 
Throughout our analysis, we  keep $\w $
large but  finite in the intermediate computations  and only  send $\w $
strictly to  infinity at the end. Therefore,  we keep explicitly terms
in the action   that  scale as   positive  powers of  $\w $ (which  look
superficially divergent) and terms that are independent of $\w $ (which are
finite). We drop terms that scale as  inverse powers of $\w $ since
they have no  chance of contributing when taking  the limit at the end
of the analysis.

In order to analyze the behaviour of the sigma model in  AdS$_5\times$S$^5$ in the non-relativistic limit, we must first compute how the various Cartan  one-forms behave under the scaling  \bref{limit}. 
Using the expressions in  Appendix \ref{appNRL}, it  is straightforward to show
that the one  forms relevant for   writing the string action \bref{totLagB} scale as
follows
\bea
\bL^\mu&=&\w \bL^{\mu(1)}+{1\over \w }\bL^{\mu(-1)}+\cdots
\nn\\
\bL^a&=&\bL^{a(0)}+\cdots
\nn\\
\bL^{m'}&=&\bL^{m'(0)}+\cdots
\nn\\
\bL^{\alpha\alpha'I}&=&\sqrt{\w }\:\bL^{\alpha\alpha'I(1/2)}+{1\over \sqrt{\w }}\:\bL^{\alpha\alpha'I(-1/2)}+\cdots,
\eea
where  $\bL^{\cdot(n)}$  is the term  scaling  as  $\w ^n$  in the  form
$\bL^\cdot$ and $\cdots$ refer to terms in the one form which do not
contribute to the action in the $\w \rightarrow \infty$ limit.
Given the expansion of the  Cartan one forms in  powers in $\w $ we  can
look  at the worldsheet action \bref{totLagB}  and  identify the terms
which are  finite and the terms  which are superficially divergent.

The finite contribution coming from the Polyakov kinetic term\footnote{An analogous analysis can be performed in the Nambu-Goto formulation. See Appendix \ref{appNGA} for a detailed analysis of $\kappa$-symmetry.} \bref{covalg} in
\bref{totLagB} is given by
\be
{\cal
L}^{(Kin)}_{fin}=\sqrt{-h}h^{ij}\left(2\eta_{\mu\nu}\bL^{\mu(1)}_i\bL_j^{\nu(-1)}+\delta_{ab}\bL_i^{a(0)}\bL_j^{b(0)}+
\delta_{m'n'}\bL_i^{m'(0)}\bL_j^{n'(0)}\right),
\label{kinfin}
\ee
while the superficially divergent contribution is given by:
\be
{\cal
L}^{(Kin)}_{div}=\w ^2\sqrt{-h}h^{ij}\eta_{\mu\nu}\bL^{\mu(1)}_i\bL_j^{\nu(1)}.
\label{kindiv}
\ee

The  finite contribution  from    the    WZ term   \bref{covalg}    in
\bref{totLagB} can be written as
\bea
{1\over    i}d(\CL_{fin}^{(WZ)}d^2\xi)&=&\;\bar  \bL^{(1/2)}\Gamma_\mu
\bL^{\mu(-1)}\tau_3\bL^{(1/2)}+
\bar \bL^{(-1/2)}\Gamma_\mu \bL^{\mu(1)}\tau_3\bL^{(-1/2)}
\nn\\
&+&  2\bar   \bL^{(1/2)}\Gamma_a   \bL^{a(0)}\tau_3\bL^{(-1/2)}+ 2\bar
\bL^{(1/2)}\Gamma_{m'} \bL^{m'(0)}\tau_3\bL^{(-1/2)},
\label{WZfin}
\eea
while the superficially divergent contribution is: 
\be
{1\over  i}d(\CL_{div}^{(WZ)}d^2\xi)=\;\w ^2\:\bar \bL^{(1/2)}\Gamma_\mu
\bL^{\mu(1)}\tau_3\bL^{(1/2)}.
\label{WZdiv}
\ee

The final term to consider is the coupling \bref{BfieldA} corresponding to turning on
a  closed $B$-field,    which  only  leads to the  following
potentially divergent term
\be
\CL_{div}^B=\w ^2 \epsilon_{\mu\nu}\bef^{\mu(1)}_0\bef^{\nu(1)}_1,
\label{Bfielddiv}
\ee
where   we   have     used   that   $\bv^\mu=\bef^{\mu(1)}$,     where
$\bL^{\mu(1)}=\bef^{\mu(1)}+
\bL_{fermionic}^{\mu(1)}$. 

In order to give a proper definition of non-relativistic string theory in AdS$_5\times$S$^5$
 we must rewrite the
superficially  divergent terms in the action
\be
\CL_{div}=-\:{T\over 2}\left(\w^2\sqrt{-h}h^{ij}\eta_{\mu\nu}\bL^{\mu(1)}_i\bL_j^{\nu(1)}+
{2}\CL_{div}^{(WZ)}+2\w^2
\epsilon_{\mu\nu}\bef^{\mu(1)}_0\bef^{\nu(1)}_1\right)
\label{totLagdiv}
\ee
in  a way  that the  $\w \rightarrow \infty$  limit yields  a consistent
worldsheet theory, which gives a proper definition of the theory. We will now show that there is a 
rich  interplay between the various terms in \bref{totLagdiv} such that when combined, conspire to yield a well defined worldsheet action, which serves as the definition of non-relativistic string theory in AdS$_5\times$ S$^5$.

In order to show this we first note that the following identity can be proven\footnote{Our convention is $\ep_{01}=-\ep^{01}=-1$.} 
\be
d(\w^2(\det\bL^{\mu(1)})_{fermionic}+\CL_{div}^{(WZ)})\=0,
\label{NIdentity}
\ee
where $(\det\bL^{\mu(1)})_{fermionic}$ is defined as follows:
\be
\det\bL^{\mu(1)}=\det\bef^{\mu(1)}+(\det \bL^{\mu(1)})_{fermionic}.
\label{relation} 
\ee 
This identity, which is proven in Appendix \ref{appDNG}, guarantees that the superficially divergent term of the Nambu-Goto Lagrangian is a total derivative. This divergence  is precisely cancelled by the background $B$-field. 

Formula \bref{NIdentity}  implies that up to an exact form one has:
\be
\CL_{div}^{(WZ)}=-\w^2(\det\bL^{\mu(1)})_{fermionic}.
\ee
 Therefore, if we combine the divergent piece of the WZ term with the divergent piece coming from the $B$-field,  the  Polyakov action can be written in terms of the Cartan  one forms $\bL^{\mu(1)}$
\be
\CL_{div}^{(WZ)}+\CL_{div}^{B}=-\w^2\det\bL^{\mu(1)},
\ee
where we have used that $\det\bef^{\mu(1)}=-\epsilon_{\mu\nu}\bef^{\mu(1)}_0\bef^{\nu(1)}_1$
and formula \bref{relation}.

After  these manipulations, we  have  that the superficially divergent
part of the Polyakov action \bref{totLagdiv} can be written as:
\be
\CL_{div}=-\w^2\:{T\over 2}\left(\sqrt{-h}h^{ij}\eta_{\mu\nu}\bL^{\mu(1)}_i\bL_j^{\nu(1)}-
2\det\bL^{\mu(1)}\right).
\ee
It is  straightforward  to show that these two  terms combine to complete
a perfect square 
\be
\CL_{div}=-\w^2\:{T\over 2}\sqrt{-h}h^{00}\h_{\mu\nu}\;f^\mu\;f^\nu,
\label{supdivtot}\ee
where:
\be
f^\mu\equiv \left[ \bL_0^{\mu(1)}-{\sqrt{-h}\over h_{11}}
\epsilon^{\mu\rho}\eta_{\rho\sigma}\bL_1^{\sigma(1)}-{h_{01}\over h_{11}}
\bL_1^{\mu(1)}\right].
\label{supdivtot2}
\ee
We can now rewrite this superficially divergent term in a way that the $\w \rightarrow \infty$ limit can be 
taken smoothly. The idea is 
to introduce Lagrange  multipliers $\lambda_{\mu}$ to rewrite  the 
action as follows
\be
\CL_{div}=\lambda_{\mu}\;f^\mu
+{1\over 2\sqrt{-h}h^{00}T\w ^2}\lambda_\mu\lambda^\mu,
\label{lagrange}
\ee
which reproduces \bref{supdivtot} by integrating out the $\lambda_\mu$ variables.
Once the extra variables $\lambda_\mu$ are introduced, we can
 take  the strict non-relativistic limit  $\w \rightarrow
\infty$  in \bref{lagrange} and be  left with a finite contribution:
\be
\CL^*=\lambda_{\mu}\;f^\mu.
\ee

Now that we have properly defined the superficial divergence, we can finally write the complete action for non-relativistic strings in AdS$_5\times$S$^5$
\bea
\CL&=&-\:{T\over 2}\left({\cal L}^{(Kin)}_{fin}+2\CL^{(WZ)}_{fin}\right)+\CL^*
\nn\\
&=&-\:{T\over
2}\sqrt{-h}h^{ij}{\bf G}^{nr}_{ij}
-\:{T}\CL^{(WZ)}_{fin}+
\lambda_{\mu}\left[ \bL_0^{\mu(1)}-{\sqrt{-h}\over h_{11}}\epsilon^{\mu\rho}
\eta_{\rho\sigma}\bL_1^{\sigma(1)}-{h_{01}\over h_{11}}\bL_1^{\mu(1)}\right],
\nn\\
\label{fin}
\eea
where  
\be
{\bf G}^{nr}_{ij}\= \eta_{\mu\nu}(\bL^{\mu(1)}_{i}\bL_j^{\nu(-1)}+\bL^{\mu(1)}_j\bL_i^{\nu(-1)})+
\delta_{ab}\bL_i^{a(0)}\bL_j^{b(0)}+
\delta_{m'n'}\bL_i^{m'(0)}\bL_j^{n'(0)} 
\label{imetnr}
\ee
and the   expression for   $\CL^{(WZ)}_{fin}$  is   determined  by
integration   of the three  form   \bref{WZfin}. 
We can now ascribe the variables $\lambda_\mu$ with a physical 
interpretation; they are linearly related
 to the momentum along the longitudinal 
directions, as can be  seen from \bref{fin}.

The gauge symmetries of the action\footnote{The structure of the Polyakov form of the Lagrangian \bref{fin} for the bosonic string in flat space 
has been discussed in \cite{Gomis:2004ht}, where it was derived from the Nambu-Goto action.}
are generated by worldsheet diffeomorphisms, Weyl transformations and $\kappa$-transformations. The existence of these symmetries can be understood as being a consequence of the existence of the corresponding symmetries of the parent relativistic theory. The limit we have found maps a symmetry of the parent theory to a corresponding symmetry of the non-relativistic theory.\footnote{Appendix \ref{appNGA}  shows that 
the symmetries of the parent theory have a counterpart in the theory after the limit.}

The action \bref{fin} is also invariant under a supersymmetric version of the Newton-Hooke group with thirty two supersymmetries! The Newton-Hooke group is precisely 
the AdS analog of the  
 Galilean group for flat space. It can be obtained by an In\"on\"u-Wigner contraction of the $SO(2,4)$ AdS symmetry in the limit in which the speed of light is sent to infinity, which is precisely how the Galilean group can be obtained from the Poincare symmetry group of flat space. The only difference 
between the Newton-Hooke group and the 
 Galilean  group is that in the 
latter the Hamiltonian commutes with translations, while in the former one 
has $[H,P_a]=i/R_0^2\, K_a$ 
 \cite{BacryLL}, where $K_a$ generate boosts between inertial frames. The commutation relations satisfied by the generators of the supersymmetric Newton-Hooke algebra can be found in Appendix \ref{appNRalg}, which can be obtained by an In\"on\"u-Wigner contraction of the $SU(2,2|4)$ algebra.

The non-relativistic Polyakov action \bref{fin}  is invariant under diffeomorphisms, albeit it is not manifest from the form of the action. 
We can also write down the action of non-relativistic string theory in AdS$_5\times$S$^5$ 
in the Nambu-Goto formulation, where invariance under diffeomorphisms is manifest. This can be done by using the equation of motion of 
$\lambda_\mu$. Using it we get:
\be
h_{ij}\propto{{\bf L}_i}^{\mu(1)}{{\bf L}_j}^{\nu(1)}\h_{\mu\nu}\equiv{\bf g}_{ij}.
\ee
Plugging this into \bref{fin} one derives the Nambu-Goto  Lagrangian of non-relativistic string theory in AdS$_5\times$S$^5$
\bea
\CL&=&
-\:{T\over
2}\sqrt{-{\bf g}}{\bf g}^{ij}{\bf G}^{nr}_{ij}
-\:{T}\CL^{(WZ)}_{fin},
\label{fincovng}
\eea
where ${\bf G}^{nr}_{ij}$ is the induced metric on the worldsheet \bref{imetnr}.
This is the Lagrangian of non-relativistic strings in AdS$_5\times$S$^5$ in the Nambu-Goto formulation.

From the action we can derive the   constraints associated with worldsheet 
diffeomorphisms.
Using the   mechanical momenta
\be
\8{p}^+_\mu=\frac{\pa \CL^{(NG)}}{\pa {\bL}_0^{\mu(1)}},\qquad 
\8{p}_a=\frac{\pa \CL^{(NG)}}{\pa {\bL}_0^{a(0)}}, \qquad 
\8{p}_{m'}=\frac{\pa \CL^{(NG)}}{\pa {\bL}_0^{m'(0)}}
\ee
and
\be
\8{p}^-_\mu=\frac{\pa \CL^{(NG)}}{\pa {\bL}_0^{\mu(-1)}}=T\ep_{\mu\nu}
{\bL}_1^{\mu(1)}
\ee
we can derive the following identities:
 \bea
H_{nr}&\equiv&
\8{p}^+_\mu \vep^{\mu\rho}\h_{\rho\s} \bL_1^{\s(1)}+
\8{p}^-_\mu \vep^{\mu\rho}\h_{\rho\s} \bL_1^{\s(-1)}+
2T \h_{\mu\nu} {\bL_1}^{\mu(1)}{\bL_1}^{\nu(-1)}
\nn\\&&+
{1\over 2T}\left[\delta^{ab}\8{p}_a\8{p}_b+\delta^{m'n'}\8{p}_{m'}\8{p}_{n'}+
T^2\left(\delta_{ab}
\bL^{a(0)}_1\bL^{b(0)}_1+\delta_{m'n'}\bL^{m'(0)}_1\bL^{n'(0)}_1\right)\right]
= 0,
\nn\\ \label{constNR}
\\
T_{nr}&\equiv&{\8p}^+_\mu {\bL_1}^{\mu(1)}+{\8p}^-_\mu {\bL_1}^{\mu(-1)}+
{\8p}_a {\bL_1}^{a(0)}+{\8p}_{m'} {\bL_1}^{m'(0)}=0.
\label{constNRT}\eea
They are diffeomorphism constraints when written in terms of the canonical 
momenta.  
They can also be 
obtained from  the corresponding relativistic ones  in \bref{constWZ1} by taking 
the non-relativistic limit.
We note that the constraint $H_{nr}$ is now {\it linear} in the energy, which is what one expects from a non-relativistic theory.

\vs

\section{Gauge fixed action}

In this section we find that the non-relativistic string action becomes free by suitably fixing all the gauge symmetries. For explicit computations, it is convenient to choose the conformal gauge 
$\sqrt{-h}h^{ij}=\eta^{ij}$.  Then the Polyakov action \bref{fin} simplifies to:
\be
\CL=-\:{T\over
2}\eta^{ij}{\bf G}^{nr}_{ij}
-\:{T}\CL^{(WZ)}_{fin}+
\lambda_{\mu}(\bL_0^{\mu(1)}-\epsilon^{\mu\rho}\eta_{\rho\sigma}\bL_1^{\sigma(1)}).
\label{fingf}
\ee
In the gauge fixed form, the term in \bref{fingf} involving $\lambda_\mu$ 
is the AdS$_5\times$S$^5$ analog of the $(\beta,\gamma)$ system introduced in \cite{Gomis:2000bd}
to describe non-relativistic string theory in flat space. In fact, in the $R_0\to\infty$ limit,  we recover the 
result for the non-relativistic superstring in the flat space \cite{Gomis:2004pw}.

In order to completely define the theory in the conformal gauge,  we must supplement the action with
 the Virasoro constraints, which are given by 
\bea
\lambda_\mu\ep^{\mu\nu}\h_{\nu\rho}\bL_1^{\rho(1)}+\frac{T}2
({\bf G}^{nr}_{00}+{\bf G}^{nr}_{11})&=&0
\nn\\
\lambda_\mu\bL_1^{\mu(1)}+T{\bf G}^{nr}_{01}&=& 0,
\label{constnon}
\eea
where ${\bf G}^{nr}_{ij}$ is defined in \bref{imetnr}.

\vs

A   simplification of the action can be achieved by fixing a gauge for $\kappa$-symmetry. In   Appendix  \ref{appNGA} we show that we can fix $\kappa$-symmetry by making the following gauge choice:
\be
\theta_-=0.
\ee
By fixing this gauge,  the Lagrangian simplifies dramatically since many terms 
  automatically   vanish, as can be seen by looking at the expressions for the one forms in Appendix \ref{appNRL}. Moreover, we can integrate the three-form \bref{WZfin} and obtain an explicit expression for the WZ term.

By using the expressions in Appendix C we have that after fixing $\kappa$-symmetry, the 
Polyakov action 
  \bref{fingf} can be written as:
\bea
\CL^{(NR)}&=&-T\;[
\frac{{\h}^{ij}}{2}\pa_ix^{a}\pa_jx^{b}\h_{ab}+
\frac{{\bf g}_{11}-{\bf g}_{00}}{2\,R_0^2}x^{a}x^{b}\delta_{ab}
+\frac{{\h}^{ij}}{2}\pa_ix^{m'}\pa_jx^{n'}\delta_{m'n'}
\nn\\&&
\-i\Tb_+ {\Gamma}^{{\mu}}(\h_{\mu\nu}\h^{ij}{{\bf v}_{j}}^{\nu}
){\bf D}_i\T_+\-i\Tb_+ 
{\Gamma}^{{\mu}}(\det{{{\bf v}_{j}}^{\nu}}){{\bf v}_{\mu}}^{i}{\bf D}_i\T_+]
\nn\\&&
\+\lambda_\mu (\bv_0^{\mu}-\epsilon^{\mu\rho}\eta_{\rho\nu}\bv_1^{\nu}).
\label{totNR10}
\eea
${\bf g}_{ij}=\h_{\mu\nu}{\bv_i}^\mu{\bv_j}^\nu$ is the metric of AdS$_2$ 
spanned  
by the longitudinal coordinates  written in terms of the worldsheet coordinates $\xi^i$. The Lagrangian has two fermionic pieces arising respectively from the 
Kinetic  and WZ term. 
The covariant derivative ${\bf D}_i\T_+$ is given  by
\be 
\bD\T_+=(d+  \frac{{1}}{2{R_0}}\bv^{\mu}\Gamma_{\mu}\s_1\tau_2 +
\frac{1}{4}\bw^{\mu\nu}\Gamma_{\mu\nu})\T_+,
\label{derivada}
\ee
where in the coordinates we are using 
\be
\bv^\mu=(dx^0,dx^1\cos\frac{x^0}{{R_0}}),\qquad \bw^{01}=
-\frac{dx^1}{{R_0}}\;\sin\frac{x^0}{{R_0}}.
\label{vielll}
\ee
In the action \bref{totNR10}, ${\bv_\mu}^i$ is the inverse of ${\bv_i}^\mu$.

The equation of motion for the Lagrange multipliers $\lambda_\mu$ is given by
\be
\bv_0^{\mu}-\epsilon^{\mu\rho}\eta_{\rho\nu}\bv_1^{\nu}=0,
\ee
which implies that the induced metric ${\bf g}_{ij}$ along the longitudinal 
coordinates is conformally flat:
\bea
{\bf g}_{00}+{\bf g}_{11}&=0,
\nn\\
{\bf g}_{01}&=0.
\label{2Dconf}
\eea
Since we are using the non-relativistic string Lagrangian in the gauge fixed form,
we must still impose the corresponding
 Virasoro conditions \bref{constnon}. 
This guarantees equivalence with the diffeomorphism invariant theory.

We can also work with the reparametrization invariant Nambu-Goto Lagrangian \bref{fincovng} where  
$\kappa$-symmetry is gauge fixed using $\theta_-=0$. From \bref{fincovng} one obtains:
\bea
\CL^{(NR)}&=&-T{\sqrt{-\det {\bf g}}}\;[
\frac{{\bf g}^{ij}}{2}\pa_ix^{a}\pa_jx^{b}\h_{ab}+
\frac{1}{R_0^2}x^{a}x^{b}\delta_{ab}\+\frac{{\bf g}^{ij}}{2}\pa_ix^{m'}\pa_jx^{n'}\delta_{m'n'}
\nn\\&&
\-2i\;\Tb_+ {\Gamma}^{{\mu}}{{\bf v}_{\mu}}^{i}{\bf D}_i\T_+].
\label{totNR1}
\eea
We note  that in the 
$R_0\to\infty$ limit,  we recover the 
result for the non-relativistic superstring in  flat space \cite{Gomis:2004pw}.

\vs

The Lagrangian \bref{totNR1} is interacting since the longitudinal scalars 
$x^\mu(\xi)$ 
are coupled to the transverse scalars $x^a(\xi)$,   $x^{m'}(\xi)$ and the dynamical fermions 
{$\theta_+(\xi)$  
via the induced AdS$_2$metric ${\bf g}_{ij}$.
 A further simplification occurs if we fix worldsheet diffeomorphisms by choosing the static gauge:
\be
x^\mu(\xi)=\xi^\mu.
\ee 
In this gauge \bref{totNR1}  becomes a free field Lagrangian. The theory describes a collection of scalars and fermions propagating in AdS$_2$. More precisely, 
we have     five  bosonic massless fields $x^{m'}$, three massive bosonic 
fields $x^a$ with $m^2=2/R_0^2$ and sixteen massive fermions $\T_+$ with $m^2=1/R_0^2$. This same two dimensional field theory in AdS$_2$ has been studied in the past in 
\cite{Bardeen:1984hm}\cite{Sakai:1984fg}\cite{Sakai:1984vm}\cite{Drukker:2000ep}.

Just as in relativistic string theory, once $\kappa$-symmetry is fixed, 
sixteen of the supersymmetries are linearly realized while the other sixteen 
are 
non-linearly realized. 
The non-linearly realized supersymmetries are generated 
by $\epsilon_+$, which induce the following transformations:
\be
\D\T_+=K\ep_+ , \qquad \D x^\mu = \D x^a=\D x^{m'}=0,
\ee
where 
\be
K=e^{-\frac{\Gamma_{0}{\s_1\tau_2}x^0}{2R_0}}e^{-\frac{\Gamma_{1}{\s_1\tau_2}x^1}{2R_0}},
\ee 
and satisfies 
\be
{\bf D}\;K\= 0.
\label{DK}
\ee
The linearly realized supersymmetries are induced by $\epsilon_-$,
\footnote{The relation between 
 the relativistic susy parameters $\epsilon$ and the non-relativistic ones is
  given by 
$\epsilon=\sqrt{\w}\epsilon_-+\frac{1}{\sqrt{\w}}\epsilon_+$.}
 which generate the following transformations:
\bea
\D x^a&=&
          -2i \Tb_+\Gamma^{a} K\ep_- ,\qquad 
\D x^{m'}=
-2i \Tb_+\Gamma^{m'} K\ep_- ,\qquad\D x^\mu=0
\\
\D\T_+&=&\frac12(\Gamma^\mu{e_\mu}^j)
(\Gamma_a\pa_jx^a+\Gamma_{m'}\pa_jx^{m'})K\ep_-
-\frac{1}{2R_0}({\Gamma_{a}x^a}-{\Gamma_{m'}x^{m'}})\s_1\tau_2 K\ep_- .
\nn\\&& \eea
Geometrically, $K\epsilon_\pm$ are the Killing spinors of the AdS$_2$ field theory in the basis of vielbeins given in \bref{vielll}.

To summarize, we have shown that the worldsheet action of non-relativistic 
string theory
in AdS$_5\times$S$^5$ reduces to a free field theory in AdS$_2$ with maximal 
supersymmetry.
Given the explicit form of the local action it is a very 
interesting problem
to study the spectrum and interactions of non-relativistic strings 
\cite{workin}.


 \section{T-dual description of non-relativistic strings}
 
 In  section $4$ we have  found  that string    propagation in  the  
non-relativistic limit \bref{fin}  is  regular
even though the AdS$_5\times$S$^5$ 
background  becomes  singular   in  the   limit.  In
particular, we found a way of rewriting the worldsheet action  such
   that the result   was  finite.  This can be accomplished because 
of  a precise cancellation between the divergent contributions from the 
Kinetic and WZ terms with the $B$-field contribution. 
The crucial   ingredient  in
 accomplishing   this was that    the divergence originating  from the
 singular  background           
\be
 ds^2=(\w ^2+\hat{\rho}^2)\left[-(dx^0)^2+\cos^2\left({x^0\over
 R_0}\right)(dx^1)^2\right]+dx^adx^a+dx^{m'}dx^{m'},
\label{singmetricagain}
\ee
 is precisely compensated  by a divergence  in the  background B field
 \be      B=\w ^2\cos\left({x^0\over  R_0}\right)dx^0\wedge        dx^1,
 \label{singB}
\ee
where $\hat{\rho}=\sqrt{x_ax^a}/R_0$.

 Just like for the  non-relativistic limit in  flat space, it is quite
 useful      to   analyze    the     fate     of     the    background
 \bref{singmetricagain},\bref{singB}  under  T-duality. We have chosen
 the parametrization of  the coset in \bref{coset} in  such a way that
 the resulting     metric \bref{metric} exhibits a   manifest   space-like  isometry
 generated   by    the  Killing  vector   $V={\partial \over  \partial
 x^1}$.  Let's compactify the  coordinate $x^1$ on  a circle of radius
 $L$.  
 Once    we   have a  finite  space-like      circle,  we  can perform a
 T-duality. The T-dual  background can be  computed by using the usual
 T-duality rules \cite{Buscher:1987sk}\cite{Bergshoeff:1995as}
 \bea 
G_{11}&=&{1\over g_{11}}\rightarrow 0
 \nn\\ G_{00}&=&g_{00}-{{g_{01}^2-B_{01}^2}\over{  g_{11}}}\rightarrow
 -2\hat{\rho}^2
\nn\\
G_{01}&=&{B_{01}\over g_{11}}\rightarrow {1\over \cos  \left({x^0\over
R_0}\right)}  \label{Tdual} \eea  while   the   rest of   the   metric
components are left  invariant. We note that the coordinate  $x^1$ has now become a
null coordinate   which  is   compactified on  a    circle  of  radius
$\alpha^\prime/L$.   Note that  after   T-duality   the background  is
regular!

By  making the  change  of  coordinates
\be
\sin\left(x^0\over R_0\right)=\tanh{\left(\t x^0\over R_0\right)},
\ee
 we find that   the T-dual of the
non-relativistic  AdS$_5\times$S$^5$ background  is the    product of a  five-dimensional
pp-wave   times  R$^5$  where  the   pp-wave  has   a   compact   null
coordinate. The metric of the T-dual background is:

\bea 
ds^2&=&2d\t    x^0dx^1-\frac{2x^ax^a}{R_0^2\cosh^2{\left(\t   x^0\over
R_0\right)}}d\t x^0d\t x^0+dx^adx^a+dx^{m'}dx^{m'}.
\eea
The background also contains a null RR flux and a dilaton giving rise to a solution of the Type IIA supergravity equations of motion. The geometrical relation between time dependent pp-waves  and
the Newton-Hooke was explored in \cite{Gibbons:2003rv}.

We have shown that the T-dual description of non-relativistic string theory in AdS$_5\times$S$^5$ is given by a time dependent pp-wave with the null coordinate compactified. Therefore, quantization of
non-relativistic strings in AdS$_5\times$S$^5$  is T-dual to the
discrete light-cone quantization (DLCQ) of the pp-wave. 
The result of this section generalizes the connection made between the quantization of 
 non-relativistic strings in flat space and  the DLCQ description of flat space \cite{Gomis:2000bd}\cite{Danielsson:2000gi}. We note that in the case considered in this paper the T-dual geometry is a different one from that of the parent theory, which we expect is a generic phenomena.

 It would be interesting to analyze string theory in this time dependent pp-wave, which can be studied by fixing the light-cone gauge.
 
\section{Comments}

In this paper  we have derived the worldsheet theory describing a sector of string theory in AdS$_5\times$S$^5$; the non-relativistic sector. In brief, we have shown that once all the gauge symmetries are fixed, that the worldsheet theory describing this sector  becomes a supersymmetric theory of free bosons and fermions propagating in an AdS$_2$ background. This opens the prospect of studying \cite{workin} the
spectrum and interactions of this worldsheet theory  to probe holography  in a new stringy regime.

One crucial aspect in  isolating  non-relativistic strings was  to consider a metric on AdS$_5$ that made manifest a space-like isometry, along which the non-relativistic strings could wind. In these coordinates, however, the metric is not static. Furthermore, the circle that the non-relativistic strings wind collapses both in the past and in the future\footnote{At $X^0=\pm{\pi \over 2}R$ in the coordinates of \bref{metric}.}. One should be able to follow the time evolution of the wound strings using the explicit worldsheet theory that we have found. It would be interesting to analyze whether the T-dual, time dependent pp-wave can shed light on this. This set-up could provide an interesting setting in which to study time dependence in string theory using the holographic dual description.

It is of great interest to fully quantize the worldsheet action and obtain the physics of both non-relativistic closed strings as well as that of open strings on D-branes. The reason that we can consider both types of strings is that  we have performed a local worldsheet analysis, without committing to specific boundary conditions on the worldsheet. There are two types of D-branes that can be considered: longitudinal and transverse to the $B$-field. Quantizing open strings on a longitudinal D-brane leads to an AdS version of non-commutative open string theory (NCOS) \cite{Seiberg:2000ms}\cite{Gopakumar:2000na}. In order to get a non-trivial open string spectrum, there is no need of a spatial circle, which is required to obtain closed non-relativistic strings. The transverse D-branes have a non-trivial spectrum only if the open strings can wind and have a non-relativistic spectrum \cite{Danielsson:2000mu}, just like the closed strings. Quantizing the model to obtain the spectrum of open and closed strings  requires understanding the proper boundary conditions of the worldsheet fields on AdS$_2$ \cite{workin}. 

 It is natural -- and very interesting -- to inquire what is the dual field theory realization of the non-relativistic string sector.  In this section we make some preliminary remarks about the possible field theory interpretation. 

The first thing that one must do  to understand the field theory description of this sector is to identify the conformal boundary of \bref{metric}. 
It is straightforward to show that the metric in the boundary is that of AdS$_2\times$S$^2$. It is given by:
\be
ds^2=-(dX^0)^2+\cos^2\left(X^0\over R\right) (dX^1)^2+d\Omega_2.
\ee
The first observation is that this metric is conformally flat and is, therefore, a consistent background geometry on which to study ${\cal N}=4$ SYM. 

Therefore, the physics of non-relativistic strings in AdS$_5\times$S$^5$ is captured by ${\cal N}=4$ SYM  on AdS$_2\times$S$^2$, but with a time dependent metric. To further isolate the physics of non-relativistic strings, we must take the limit in \bref{limit}. From the point of view of the field theory on the boundary, this has the effect of shrinking the size of the S$^2$ relative to AdS$_2$. Therefore, the ${\cal N}=4$ theory effectively reduces to a 1+1 dimensional field theory living in AdS$_2$. It is this reduction of ${\cal N}=4$ SYM that encodes the physics of non-relativistic strings in AdS$_5\times$S$^5$. It would be interesting to understand the integrability structure of this theory and its relation with integrability on the worldsheet. 

In recent years, we have learned that in some situations, the dual gauge reproduces the worldsheet theory in the appropriate sector. In our case there  
is a tantalizing connection between the AdS$_2$ field theory living on the worldsheet and the 1+1 dimensional field theory that arises by taking the corresponding limit of ${\cal N}=4$ SYM. It would be very interesting to understand this relation in more detail \cite{workin}.

\newpage

\vskip 4mm

{\bf Acknowledgments}

We acknowledge discussions with Jan  Brugues, Juan Luis Ma\~nes, Rob Myers, 
Filippo
Passerini, Soo-Jong Rey, Joan Sim\'on, Arkady Tseytlin, Paul Townsend and Toine Van Proeyen.

Joaquim Gomis acknowledges the Francqui Foundation for the 
Interuniversity International Francqui chair in Belgium 
awarded to him, as well as the 
warm hospitality at the University of Leuven. 
This work is partially supported by MRTN-CT-2004-005104, 
MYCT FPA 2004-04582-C02-01, CIRIT GC 2001 SGR-00065.

\vskip 4mm

\appendix 
 
\section{Invariant one forms}\label{appLIF}

The superstring action in the AdS$_5\times$S$^5$ background is formulated using the coset superspace
SU(2,2$|$4)/((SO(4,1)$\times$ SO(5)) \cite{Metsaev:1998it}.  
We will  mainly follow the notation used in that paper. 
The vector index for AdS$_5$ is $m=0,1,2,3,4$ and that for $S^5$ is 
$m'=1',2',3',4',5'$ with the 
flat metric $\h_{mn}=diag(-;++++)$ and $\delta_{m'n'}=diag(+++++)$ . 
 The 10D gamma matrices are $\Gamma^m=\gamma^m\otimes 1 \otimes \s_1\;$ for the
AdS$_5$ part and 
 $\Gamma^{m'}=1\otimes \gamma^{m'} \otimes \s_2$  for S$^5$. 
The charge conjugation matrix is $\CC=CC'(i\s_2)$ and $\bar\theta\equiv \theta^T \CC$.
The chirality matrix is defined by 
$\Gamma_0...\Gamma_4\Gamma_{1'}...\Gamma_{5'}=\s_3$ and 
$\T$ has the  plus chirality $\s_3\T=\T$. 
 The AdS$_5$ $\gamma$-matrix  indices are $\A=1,2,3,4,\;$ and $\A'=1,2,3,4$ for S$^5$.  
$A=1,2 $ are the indices of the $\sigma-$matrices 
and are often abbreviated.
$I=1,2 $ are the SL(2,R) indices on which the $\tau-$matrices act.

Given the coset parametrization in \bref{coset}
\be
g\=e^{iP_1 X^1}e^{iP_0 X^0}e^{iP_a X^a}e^{iP_{m'} X^{m'}}e^{iQ_{\A\A'I}\T^{\A\A'I}},
\label{cosetapp}
\ee
we can readily compute $\Omega=-ig^{-1}dg$ \bref{forms} using the $SU(2,2|4)$ algebra.  
The forms associated with the bosonic generators of $SU(2,2|4)$ are given by
\bea
\matrix{
\bL^m\=e^m+\Tb\DELTA^{m\sharp}C_UD\T,
&\bL^{mn}=w^{mn}+\frac{1}{R}\Tb\DELTA^{mn}C_UD\T,
\cr 
\bL^{m'}\=e^{m'}+\Tb\DELTA^{m'\sharp'}C_UD\T,
&
\bL^{m'n'}=w^{m'n'}+\frac{1}{R}\Tb\DELTA^{m'n'}C_UD\T,
}\label{LIF}
\eea
while the form associated to the supersymmetry generators is 
\be
\bL\=S_UD\T.
\ee
Here we also use SO(4,2) and SO(6) $\Gamma$ matrix  tensors $\Gamma_{MN}, \DELTA^{MN},
(M=0,1,2,3,4,\sharp)$
and  $\Gamma_{M'N'}, \DELTA^{M'N'},(M'=1',2',3',4',5',\sharp')$ defined by 
\bea
\matrix{\Gamma_{mn}=\gamma_{mn},& \Gamma_{m\sharp}=\gamma_{m}\,\tau_2&
{\DELTA}^{mn}=\gamma^{mn}\,\tau_2\,\s_2,&{\DELTA}^{m\sharp}=\gamma^{m}(-\s_2),\cr
\Gamma_{m'n'}=\gamma_{m'n'},&
\Gamma_{m'\sharp'}=\gamma_{m'}(i\tau_2),&
{\DELTA}^{m'n'}=\gamma^{m'n'}(-\tau_2\,\s_2),&
{\DELTA}^{m'\sharp'}=\gamma^{m'}(-i\s_2).}
\nn\\
\eea
$e^{m,m'}$ are the bosonic 
{vielbeins} that determine the AdS$_5\times$ S$^5$ metric and $w^{\cdot\cdot}$ are the usual bosonic spin connections associated with the stability 
group of the coset. 
$D\T$ is the covariant derivative 
\be
D\T\=d\T+ \frac{1}{4}w^{mn}\;\Gamma_{mn}\T+ \frac{1}{2R}e^{m}\;\Gamma_{m\sharp}\T+\frac{1}{4}w^{m'n'}\;\Gamma_{m'n'}\T+\frac{1}{2R}e^{m'}\;\Gamma_{m'\sharp'}\T
\ee
satisfying  $\;D^2\T=0\;$ and  $C_U$ and $S_U$ are matrices defined by 
\be 
S_U\=\frac{\sinh {U}}{{U}},\quad
C_U\=2\frac{\cosh{U}-1}{{U^2}},
\label{SCU}
\ee
\be
U^2\equiv\frac{1}{R}\left(\frac{1}{2}(\Gamma_{mn}\T)(\Tb{\DELTA}^{mn})+
(\Gamma_{m\sharp}\T)(\Tb{\DELTA}^{m\sharp})
+\frac{1}{2}(\Gamma_{m'n'}\T)(\Tb{\DELTA}^{m'n'})+(\Gamma_{m'\sharp'}\T)
(\Tb{\DELTA}^{m'\sharp'})\right),
\label{U2}\ee
where $R$ is the radius of AdS$_5$ and  S$^5$.

The expressions for $e^m$ and $w^{mn}$ are given by
$(\rho=\sqrt{X^a X_a}/R),$
\bea
e^0&=&dX^0\;\cosh{{\rho}}\;,\qquad e^1\=dX^1\;\cosh{\rho}\;\cos\frac{X^0}{R}\;,
\label{e0}\\
e^a&=&
dX^a+dX^b
({\h_b}^a-\frac{X_bX^a}{\rho^2 R^2}) \left( 
\frac{\sinh{\rho}}{{\rho}}-1\right) ,
\label{ea}\\
 w^{01}&=&-\frac{dX^1}{R}\;\sin\frac{X^0}{R},
\label{w01}\\
w^{0a}&=&\frac{X^ad X^0}{\rho R^2}\;\sinh{{\rho}},\qquad
w^{1a}\=\frac{X^ad X^1}{\rho R^2}\;\sinh{{\rho}}\;\cos\frac{X^0}{R},
\label{w0a}\\
 w^{ab}&=&
\frac{dX^{a}X^{b}-dX^{b}X^{a}}{\rho^2R^2}\left( \cosh{\rho}-1\right).
\label{wab}\eea
while $e^{m'}$ and $w^{m'n'}$ are given by
$(r=\sqrt{X^{m'} X_{m'}}/R)$,
\bea
e^{m'}&=&dX^{m'}+ 
dX^{n'}({\h_{n'}}^{m'}-\frac{X_{n'}X^{m'}}{{r}^2R^2})\left( 
\frac{\sin{r}}{r}-1\right),
\\
w^{m'n'}&=&
\frac{dX^{m'}X^{n'}-dX^{n'}X^{m'}}{{r}^2R^2}\; (\cos{r}-1).
\eea
\vs

\section{Embedding of AdS }\label{AppEmbed}

The AdS$_5$ geometry is embedded in a flat $6$ dimensional space with the 
$SO(4,2)$  invariant metric
$\h_{MN}=(-;++++;-)$
\be
ds^2=\h_{MN}du^Mdu^N.
\ee 
It is a hyperboloid satisfying:
\be
\h_{NM}u^Mu^N\=-{u^0}^2+{u^1}^2+(\sum_{a=2}^4{u^a}^2)-{u^\sharp}^2\=-R^2. 
\label{umun}
\ee
For the parametrization of the coset \bref{coset}   
\be
g\=e^{iP_1 X^1}e^{iP_0 X^0}e^{iP_a X^a},
\label{cosetads}
\ee
the left invariant one forms are given in \bref{e0}-\bref{wab}.
The relation between the 6D coordinates $u^M$ and the 5D coordinates $X^m$ is:
\bea
u^M&=&\pmatrix{u^0 \cr u^1 \cr u^a \cr u^5}\=R\pmatrix{
\cosh{\rho}\sin{\frac{X^0}{R}} \cr 
\cosh{\rho}\cos{\frac{X^0}{R}}\sinh{\frac{X^1}{R}} \cr
\sinh{\rho}\;\frac{X^a}{R\rho} \cr  
\cosh{\rho}\cos{\frac{X^0}{R}}\cosh{\frac{X^1}{R}}}.
\label{AdScoo}
\eea
The parametrization of \bref{AdScoo} is not global as the AdS$_2$ part of
the coset parametrization \bref{cosetads}, $e^{iP_1 X^1}e^{iP_0 X^0}$,
is not a global one. The conformal boundary is AdS$_2\times$S$^2$, where
the dual field theory lives. 


\vs

In the NR limit \bref{limit} we use $X^\mu= \w x^\mu,\;R \to \w R_0$.
In this limit the AdS$_5$ hyperboloid becomes
\bea
u^M
&\too&\w R_0\pmatrix{
\sin{\frac{x^0}{R_0}} \cr 
\cos{\frac{x^0}{R_0}}\sinh{\frac{x^1}{R_0}} \cr
\frac1{\w}\,\frac{x^a}{ R_0} \cr  
\cos{\frac{x^0}{R_0}}\cosh{\frac{x^1}{R_0}}}\too \w R_0\pmatrix{
\sin{\frac{x^0}{R_0}} \cr 
\cos{\frac{x^0}{R_0}}\sinh{\frac{x^1}{R_0}} \cr {\bf 0}\cr  
\cos{\frac{x^0}{R_0}}\cosh{\frac{x^1}{R_0}}}.
\label{rescalecoord}
\eea
Using \bref{umun} with the 
 renormalized coordinates $\8u^M=\frac{u^M}{\w}$, \bref{rescalecoord} becomes the 
   parametrization  of AdS$_2$ 
with radius $R_0$.  

Although these are not global coordinates, there is an almost everywhere space-like
Killing vector $V=\pa_{x^1}$.  In these coordinates we can make the following identification

\be
x^1\simeq x^1+L.
\ee
The metric of the compactified space is singular at the time
boundaries of the chart $x^0=\pm \frac{\pi}{2} R_0$.

\section{Non-relativistic limit} \label{appNRL}

In the non-relativistic limit \bref{limit} 
the coordinates are scaled as
\be
X^\mu\=\w x^\mu, \qquad \theta\=\sqrt{{\w} }\:\theta_-+{1\over \sqrt {\w} }\:
\theta_+,\qquad 
B_{\mu\nu}\=\epsilon_{\mu\nu}, \qquad R\={\w} R_0.
\label{limita}
\ee
$\T_\pm$ are defined using  the projection operators $\bP_\pm$ ,
\be
\bP_\pm\T_\pm=\T_\pm,\qquad
\bP_\pm=\frac12(1\pm\Gamma_*),\qquad \Gamma_*\equiv\Gamma_0\Gamma_1\tau_3. 
\label{ppm}
\ee
In this scaling, (in the following, we only write  terms with negative powers of ${\w} $ which are  relevant in the limit ${\w} \to\infty$)  
\be
e^0\= {\w } (1+\frac{\hat\rho^2}{2{\w} ^2})dx^0,\qquad 
e^1\= {\w } (1+\frac{\hat\rho^2}{2{\w} ^2})dx^1\cos\frac{x^0}{{R_0}},\qquad 
e^a\=dx^a.
\ee
Since the leading term in $U^2$ in \bref{U2} is of order ${\w} ^0$, the 
trigonometric 
factors $C_U$ and $S_U$ in \bref{SCU} are an infinite series at the order 
${\w} ^0$
(which is actually truncated due to the fermions). 
However in the $\kappa$-symmetry  gauge choice $\T_-\=0$ in \bref{kappatheta} 
the leading terms in $U^2$ vanish and
\bea
\bP_\pm C_U\bP_\pm=1
+\frac{1}{\w^2}(C_U)_{\pm\pm}^{(-2)},\qquad 
\bP_\pm C_U\bP_\mp\=\frac{1}{\w}
(C_U)_{\pm\mp}^{(-1)}
\eea
and the same for $S_U$. In this gauge it is also shown that the leading ${\w} ^{\frac12}$ term of $D\T$ disappears and starts with  a ${\w} ^{-\frac12}$ term
\be
\bP_+(D\T)=\frac{1}{\sqrt{{\w} }}\bD\T_+,\qquad \bP_-(D\T)=O(\frac{1}{{{\w} }^{3/2}}),
\ee
where
\be 
\bD\T_+=(d+  \frac{{1}}{2{R_0}}\bv^{\mu}\gamma_{\mu}\tau_2 +
\frac{1}{4}\bw^{\mu\nu}\gamma_{\mu\nu})\T_+,
\ee
\be
\bv^\mu=(dx^0,dx^1\cos\frac{x^0}{{R_0}}),\qquad \bw^{01}= w^{01}=-\frac{dx^1}{{R_0}}\;\sin\frac{x^0}{{R_0}}
\ee

These  expressions all simplify drastically in the $\w\to 0$ limit.
The  Cartan one forms \bref{LIF} 
in the \NR limit in the $\T_-\=0$ gauge 
\bref{kappatheta} are  
\bea
\bL^\mu&=&{\w} \bL^{\mu(1)}+{1\over {\w} }\bL^{\mu(-1)}+\cdots
\={\w} \;\bv^\mu+{1\over {\w} }\;(\frac{\hat\rho^2}{2}\bv^\mu+\Tb_+\DELTA^{\mu\sharp}\bD\T_+)+\cdots
\nn\\
\bL^a&=&\bL^{a(0)}+\cdots\= dx^a+\cdots
\nn\\
\bL^{m'}&=&\bL^{m'(0)}+\cdots\= dx^{m'}+\cdots
\nn\\
\bP_+\bL^{\alpha\alpha'I}&=&{1\over \sqrt{{\w} }}\:\bL_+^{\alpha\alpha'I(-1/2)}+\cdots
\={1\over \sqrt{{\w} }}\;(\bD\T_+)^{\alpha\alpha'I}+\cdots,
\nn\\
\bP_-\bL^{\alpha\alpha'I}&=&{1\over {{{\w} }^{3/2}}}\:\bL_-^{\alpha\alpha'I(-3/2)}
+\cdots\=\cdots.
\eea
where $\cdots$ are terms that will not contribute to the action 
in the ${\w} \rightarrow \infty$ limit. 

\vs
\section{Divergent terms of the Lagrangian in the Nambu-Goto form}
\label{appDNG}

Here, we derive the crucial identity
\be
d(\w^2(\det\bL^{\mu(1)})_{fermionic}+\CL_{div}^{(WZ)})\=0
\ee
used in the main  text \bref{NIdentity}. Since $\det\bL^{\mu(1)}=\det\bef^{\mu(1)}+(\det\bL^{\mu(1)})_{fermionic}$ and $\det\bef^{\mu(1)}$ is trivially closed (it is a top form), \bref{NIdentity} is equivalent to 
\be
d(\w^2\det\bL^{\mu(1)}+\CL_{div}^{(WZ)})\=0.
\label{NIdentitya}
\ee
Noting that $\ep^{01}=-\ep_{01}=1$, we have 
\bea
\det (\bL^\mu_i)d^2\xi&=&-\frac12\ep_{\mu\nu}\bL^\mu\bL^\nu
\eea
Multiplying by $d$ and using the Maurer-Cartan equation \bref{MCeq} we get 
\bea
d[ \det (\bL^\mu_i)d^2\xi]&=&-\ep_{\mu\nu}(d\bL^\mu)\bL^\nu
=-\ep_{\mu\nu}(-\bL^{\mu\rho}{\bL_{\rho}}-\bL^{\mu a}{\bL_{a}}+
\bar\bL\DELTA^{\mu\sharp}\bL)\bL^\nu.
\eea
The $\w^2$ term in $d\CL^{(NG)}$ comes from the last term and satisfies

\bea
[d[-T\sqrt{-\det g}\;d^2\xi]]_{{\w^2\,term}}&=&T\ep_{\mu\nu}
(\bar\bL^{(\frac12)}_-\DELTA^{\mu\sharp}
\bL^{(\frac12)}_-)\bL^{\nu(1)}.
\label{NGw2}
\eea
Using an identity which holds from the definition of   $\T_-$
\bea
\ep_{\mu\nu}\DELTA^{\mu\sharp}\bP_-&=&\ep_{\mu\nu}\gamma^\mu(-\s_2)\bP_-
=-i\Gamma_\nu\s_3\tau_3\bP_-
\eea
the divergent piece of the Nambu-Goto term can be written as 
\bea
[d[-T\sqrt{-\det g}\;d^2\xi]]_{{\w^2\,term}}&=&
T(\bar\bL^{(\frac12)}_-(-i\Gamma_\nu \tau_3)\bL^{(\frac12)}_-)\bL^{\nu(1)}.
\label{NGw21}\eea

On the other hand the WZ action is given by \bref{covalg}
\bea
d\CL^{(WZ)}&=&-iT\;\bar\bL(\Gamma_m \bL^m+\Gamma_{m'} \bL^{m'})\tau_3
\bL.
\eea
The $\w^2$ term comes from $m=\mu$ and by isolating the $\bL_-$ term as
\be
[d\CL^{(WZ)}]_{\w^2\,term}=
-iT\,(\bar\bL^{(\frac12)}_-\Gamma_\mu \bL^{\mu(1)})\tau_3\,
\bL^{(\frac12)}_-
\label{WZw21}\ee
and it cancels with \bref{NGw21},  which implies \bref{NIdentitya}.
Therefore, the $\w^2$ piece of the non-relativistic Nambu-Goto Lagrangian is a 
total derivative.

  
\section{Nambu-Goto form of the action and $\kappa$-symmetry}\label{appNGA}

By integrating out the auxiliary metric $h_{ij}$ in the Polyakov action \bref{covalg}, we get the Nambu-Goto action:
\be
\CL\=-T\;\CL^{(NG)}\-T\;\CL^{(WZ)}.
\label{totLaga}
\ee
The first term is the Nambu-Goto Kinetic term {\bref{NGaction}}
\be
\CL^{(NG)}\=\sqrt{-\det G_{ij}},
\label{NGaction3}
\ee
with the induced metric 
\be
G_{ij}=
\bL^{m}_i\bL^{n}_j\h_{mn}+\bL^{m'}_i\bL^{n'}_j\delta_{m'n'}.
\label{NGaction1}
\ee
The second contribution is the WZ Lagrangian whose exterior derivative is given as 
an invariant three form
\bea
d(\CL^{(WZ)}d^2\xi)
&=&i\,\bar \bL(\Gamma_m \bL^m+\Gamma_{m'} \bL^{m'})\tau_3\bL.
\label{dWZaction}
\eea
$\bL^m, \bL^{m'}$ and $\bL^{\A\A'AI}$ are left invariant one forms 
constructed from the coset \noindent$\frac{SU(2,2|4)}{SO(4,1)\times
SO(5)}$, see Appendix \ref{appLIF}.

This relativistic action is  invariant under diffeomorphism and $\kappa$-symmetry
\cite{Metsaev:1998it}. 
The $\kappa$-symmetry transformations are
\be
[\D   \T]=\frac12(1-\Gamma_\kappa)\kappa \qquad {\rm  and}  \qquad 
[\D X]^{m} \= [\D X]^{m'} \=0,
\label{kappat}
\ee 
where $[\delta \T]^{\alpha\alpha'I}, [\D X]^{m}$ and $[\D X]^{m'}$  
are  $\bL^{\alpha\alpha'I}, \bL^{m}$ and $\bL^{m'}$ in which $dX^A$ is replaced by $\D X^A$ for
the   superspace  coordinates  $X^A=(X^m,X^{m'},\T)$\footnote{
In the Polyakov form of the Lagrangian the metric $h_{ij}$ is also transformed
properly \cite{Metsaev:1998it}.}.  $\Gamma_\kappa$
satisfies $ \Gamma_\kappa^2=1$ and is given by
\be
\Gamma_\kappa\equiv\frac{1}{2\sqrt{-G}}\ep^{ij}\slG_i\slG_j\tau_3,\qquad
\slG\equiv\Gamma_m \bL^m+\Gamma_{m'} \bL^{m'},
\ee
where $G$ is the determinant of the induced metric. 
The  $\kappa$-symmetry transformations $\D   X^A$ of the supercoordinates
are determined from \bref{kappat}.

Let us know find the $\kappa$-transformation for the scaled variables 
$\theta_\pm$ as an expansion in terms of $\omega$. In order to do that
we rescale the parameter $\kappa$
\be
\kappa=\sqrt{\w}\kappa_-+\frac{1}{\sqrt{\w}}\kappa_+
\ee
and we do 
the power expansion of $\Gamma_\kappa$ in powers of $\w$
\bea
\Gamma_\kappa&=&\Gamma_*+ 
\frac{\ep^{ij}}{\w\sqrt{-g}}
(\Gamma_\mu{\bL^\mu_i}^{(1)})                 (\Gamma_b{\bL^b_j}^{(0)}
+\Gamma_{m'}{\bL^{m'}_j}^{(0)} )
\tau_3+O(\w^{-2})
\eea
where $g=\det(\h_{\mu\nu}\bL_i^{\mu(1)}\bL_j^{\nu(1)})$. We have
\bea
[\D \T]_-= \kappa_-,\qquad
[\D \T]_+=-
\frac{\ep^{ij}}{2\sqrt{-g}}
(\Gamma_\mu{\bL^\mu_i}^{(1)})                 (\Gamma_b{\bL^b_j}^{(0)}
+\Gamma_{m'}{\bL^{m'}_j}^{(0)} )
\tau_3 \kappa_- +\cdots
\label{nrkappa}
\eea
Notice that $\kappa_+$ does not contribute the lowest order in
$\omega$.
 From
\bref{nrkappa}  we can see that $\T_-$  could be gauged away using the
$\kappa$-transformation since
\be
\D\T_-|_{\T_-=0}=\kappa_-.
\ee
Therefore, $\T_-$ is the gauge degree of freedom associated to $\kappa$-transformation. In this paper we choose the $\kappa$-symmetry gauge condition:
\be
\T_-\=0.
\label{kappatheta}
\ee
This gauge is also  valid for the Polyakov action. 
Especially since $\T_-=0$ is stable under 
the diffeomorphism and the Weyl symmetry, 
the choice is used in the conformal gauge of the metric.

We thus see that the non-relativistic limit guarantees that the non-relativistic string action inherits the 
gauge symmetries of the parent theory. 
The symmetries of the non-relativistic Lagrangian are a consequence of the 
symmetries of the parent relativistic action and the fact that the 
divergent term of the non-relativistic expansion of the relativistic action is 
total derivative, as we proved in Appendix \ref{appDNG}.

\vs
\vs
\section{Non-relativistic string contraction of $SU(2,2|4)$ algebra}
\label{appNRalg}

The space-time  supersymmetry algebra 
of the non-relativistic action \bref{fin}\bref{fincovng} is given by a supersymmetrization of the Newton-Hooke group. 
It can be obtained by  the non-relativistic contraction of
$SU(2,2|4)$ . Let us study this algebra
using  SO(4,2)$\times$SO(6) notations.  Letting $\8M=(M,M')$ run over SO(4,2) and SO(6) indices the algebra is
\bea
\left[M_{\8M\8N},M_{\8R\8S}\right]&=&-i\h_{\8N[\8R}M_{{\8M\8S]}}+
i\h_{\8M[\8R}M_{{\8N\8S]}},
\label{MMalg}\\
\left[Q_\pm,~M_{\8M\8N}\right]&=&\frac{i}{2}Q_\pm(\Gamma_{{\8M\8N}})
\label{QMalg}\\
\{ Q_{\A\A'I},~Q_{\B\B'J}\}& =&
-\frac{i}{R}(\CC \DELTA^{{\8M\8N}})_{{\A\A'I},{\B\B'J}}~M_{\8M\8N},
\label{QQalg}\eea
where $P_m=\frac{1}{R}M_{m\sharp},\; P_{m'}=\frac{1}{R}M_{m'\sharp'}.$
Corresponding to the rescaling \bref{limit}
we rescale the generators as 
\be
P_\mu \to  \frac{1}{\w} P_\mu ,\quad
M_{\mu a}\to \w B_{\mu a},\quad Q_-\to
\frac{1}{\sqrt{\w}}Q_-,\quad Q_+\to{\sqrt{\w}}Q_+,
\ee
where the supersymmetry  generators $Q_\pm$ are defined using the projection operator 
$\bP_\pm=\frac12(1\pm\Gamma_0\Gamma_1\tau_3).$ The non-zero bosonic $AdS_5$ commutators in the $\w\to\infty$ limit become 
\bea
\left[  P_\mu,~  M_{\nu\rho}\right]&=&-i\h_{\mu[\nu}~  P_{\rho]},\qquad
\left[  M_{\mu\nu},  B_{\rho b}\right]\=-i\h_{[\nu\rho}  B_{\mu] b},
\nn\\
\left[  P_a,  M_{cd}\right]&=&-i\h_{a[c}~  P_{d]},\qquad
\left[  B_{\mu a},  M_{cd}\right]\=-i\h_{a[c}  B_{\mu d]},
\nn\\
\left[  M_{ab},  M_{cd}\right]&=&-i(\h_{b[c}  M_{ad]}-
\h_{a[c}  M_{bd]})
\eea
and 
\bea
 \left[  P_\mu, ~ P_\nu\right]&=&-i~(\frac{1}{  R_0^2})~  M_{\mu\nu} ,\qquad
\left[  P_\mu,~  P_b\right]\=-i~(\frac{1 }{  R_0^2})~B_{\mu b},
\nn\\
\left[  P_\mu,~  B_{\nu b}\right]&=&-i~\h_{\mu\nu}~P_{b}.
\label{eq4}
\eea
The bosonic $S^5$ algebra becomes Euclidian algebra in the limit,
\bea
\left[  P_{m'},~  M_{n'\l'}\right]&=&-i\h_{m'[n'}~  P_{\l']},
\nn\\
\left[  M_{m'n'},  M_{\l'r'}\right]&=&-i(\h_{n'[\l'}  M_{m'r']}-
\h_{m'[\l'}  M_{n'r']}).
\eea
\vs

The QM algebra \bref{QMalg} becomes using
\bea
\Gamma_{\mu\sharp }\bP_\pm&=&\bP_\pm\Gamma_{\mu\sharp },\quad 
\Gamma_{a\sharp }\bP_\pm=\bP_\mp\Gamma_{a\sharp },\quad 
\Gamma_{\mu\nu}\bP_\pm=\bP_\pm\Gamma_{\mu\nu},
\quad 
\Gamma_{\mu a}\bP_\pm=\bP_\mp\Gamma_{\mu a},\label{ppmgam}
\nn\\
\Gamma_{ab}\bP_\pm&=&\bP_\pm\Gamma_{ab},\quad 
\Gamma_{m'\sharp' }\bP_\pm=\bP_\mp\Gamma_{m'\sharp' },\quad 
\Gamma_{m'n'}\bP_\pm=\bP_\pm\Gamma_{m'n'},
\label{ppmgam2}
\eea
\bea
\left[Q_\pm,~P_{\mu}\right]&=&~\frac{i}{2R_0}
Q_\pm(\Gamma_{{\mu\sharp}}),\qquad 
\left[Q_-,~P_{{a}}\right]=\frac{i}{2R_0}
Q_+(\Gamma_{{a\sharp}}),
\nn\\
\left[Q_\pm,~M_{\mu\nu}\right]&=&~\frac{i}{2}
Q_\pm(\Gamma_{{\mu\nu}}),\qquad \left[Q_-,~P_{{m'}}\right]=\frac{i}{2R_0}
Q_+(\Gamma_{{m'\sharp'}})
\nn\\
\left[Q_\pm,~M_{ab}\right]&=&~\frac{i}{2}
Q_\pm(\Gamma_{{ab}}),\qquad
\left[Q_-,~B_{{\mu a}}\right]=\frac{i}{2}
Q_+(\Gamma_{{\mu a}}),
\nn\\
\left[Q_\pm,~M_{m'n'}\right]&=&~\frac{i}{2}
Q_\pm(\Gamma_{{m'n'}}).
\eea
The QQ algebra \bref{QQalg} becomes using
\bea
&&\bP^T_\pm\;\CC~=~\CC \bP_\mp,
\nn\\
\DELTA^{\mu\sharp}\bP_\pm&=&\bP_\mp\DELTA^{\mu\sharp},\quad 
\DELTA^{a\sharp}\bP_\pm=\bP_\pm\DELTA^{a\sharp},\quad 
\DELTA^{\mu\nu}\bP_\pm=\bP_\mp\DELTA^{\mu\nu},\quad 
\DELTA^{\mu a}\bP_\pm=\bP_\pm\DELTA^{\mu a},
\nn
\\
\DELTA^{ab}\bP_\pm&=&\bP_\mp\DELTA^{ab},\quad 
\DELTA^{m'\sharp '}\bP_\pm=\bP_\pm\DELTA^{m'\sharp '},\quad 
\DELTA^{m'n'}\bP_\pm=\bP_\mp\DELTA^{m'n'},
\eea
and taking $\w\to\infty $ limit
\bea
\{ Q_{-},~Q_{-}\} &=&
-\frac{i}{R_0}[\CC (2\DELTA^{\mu\sharp}~R_0P_{\mu}+\DELTA^{\mu\nu}M_{\mu\nu}
+\DELTA^{ab}M_{ab}+\DELTA^{m'n'}M_{m'n'})\bP_-],
\nn\\
\{ Q_{+},~Q_{-}\} &=&
-\frac{2i}{R_0}[\CC (\DELTA^{a\sharp}~R_0P_{a}+\DELTA^{\mu a}B_{\mu a}
+\DELTA^{m'\sharp'}~R_0P_{m'})\bP_-].
\eea



\begin{thebibliography}{99}

\bibitem{Berenstein:2002jq}
  D.~Berenstein,  J.~M.~Maldacena and  H.~Nastase, ``Strings  in flat
  space and pp  waves from N = 4  super Yang Mills,''  JHEP {\bf 0204}
  (2002) 013 [arXiv:hep-th/0202021].  


\bibitem{Hatsuda:2002xp}
  M.~Hatsuda,  K.~Kamimura  and M.~Sakaguchi, ``From  super-AdS(5)  x
  S**5 algebra  to super-pp-wave algebra,'' Nucl.\  Phys.\ B {\bf 632}
  (2002) 114 [arXiv:hep-th/0202190].  




\bibitem{Gomis:2000bd}
J.~Gomis and H.~Ooguri, ``Non-relativistic closed string theory,'' J.\
Math.\ Phys.\ {\bf 42} (2001) 3127 [arXiv:hep-th/0009181].

\bibitem{Danielsson:2000gi}
U.~H.~Danielsson,  A.~Guijosa and   M.~Kruczenski, ``IIA/B,  wound and
wrapped,'' JHEP {\bf 0010} (2000) 020 [arXiv:hep-th/0009182].



\bibitem{Brugues:2004an}
  J.~Brugues,   T.~Curtright,  J.~Gomis   and           L.~Mezincescu,
  ``Non-relativistic strings and branes as non-linear realizations of
  Galilei 
  [arXiv:hep-th/0404175].  


\bibitem{Gomis:2004pw}
  J.~Gomis, K.~Kamimura     and    P.~K.~Townsend, ``Non-relativistic
  superbranes,''   JHEP {\bf 0411}  (2004) 051 [arXiv:hep-th/0409219].

\bibitem{workin}
work in progress





\bibitem{Metsaev:1998it}
  R.~R.~Metsaev and A.~A.~Tseytlin,  ``Type IIB superstring  action in
  AdS(5) x S(5)  background,'' Nucl.\ Phys.\  B  {\bf 533} (1998)  109
  [arXiv:hep-th/9805028].  


\bibitem{Seiberg:2000ms}
  N.~Seiberg, L.~Susskind and N.~Toumbas,
  ``Strings in background electric field, space/time noncommutativity  and a
  new noncritical string theory,''
  JHEP {\bf 0006}, 021 (2000)
  [arXiv:hep-th/0005040].

\bibitem{Gopakumar:2000na}
  R.~Gopakumar, J.~M.~Maldacena, S.~Minwalla and A.~Strominger,
  ``S-duality and noncommutative gauge theory,''
  JHEP {\bf 0006}, 036 (2000)
  [arXiv:hep-th/0005048].

\bibitem{Klebanov:2000pp}
  I.~R.~Klebanov and J.~M.~Maldacena,
  ``1+1 dimensional NCOS and its U(N) gauge theory dual,''
  Int.\ J.\ Mod.\ Phys.\ A {\bf 16}, 922 (2001)
  [Adv.\ Theor.\ Math.\ Phys.\  {\bf 4}, 283 (2000)]
  [arXiv:hep-th/0006085].
  
  
  \bibitem{Danielsson:2000mu}
  U.~H.~Danielsson, A.~Guijosa and M.~Kruczenski,
  ``Newtonian gravitons and D-brane collective coordinates in wound string
  JHEP {\bf 0103}, 041 (2001)
  [arXiv:hep-th/0012183].

  


\bibitem{Garcia:2002fa}
J.~A.~Garcia, A.~Guijosa and J.~D.~Vergara, ``A membrane action for OM
theory,'' Nucl.\ Phys.\ B {\bf 630} (2002) 178 [arXiv:hep-th/0201140].



\bibitem{BacryLL} H.~Bacry and JM.~L\'evy-Leblond, "Possible Kinematics",  J. Math. Phys.
{\bf 9} (1967), 1605.

\bibitem{Gomis:2004ht}
  J.~Gomis and F.~Passerini,
  ``Rotating solutions of non-relativistic string theory,''
  Phys.\ Lett.\ B {\bf 617}, 182 (2005)
  [arXiv:hep-th/0411195].




\bibitem{Bardeen:1984hm}
  W.~A.~Bardeen and D.~Z.~Freedman, ``On The Energy Crisis In Anti-De
  Sitter Supersymmetry,''   Nucl.\  Phys.\ B   {\bf  253}  (1985) 635.

\bibitem{Sakai:1984fg}
  N.~Sakai  and Y.~Tanii,  ``Effective  Potential In  Two-Dimensional
  Anti-De  Sitter Space,'' Nucl.\   Phys.\   B {\bf 255}  (1985)  401.
  
  \bibitem{Sakai:1984vm}
  N.~Sakai and Y.~Tanii,
  ``Supersymmetry In Two-Dimensional Anti-De Sitter Space,''
  Nucl.\ Phys.\ B {\bf 258}, 661 (1985).



\bibitem{Drukker:2000ep}
  N.~Drukker, D.~J.~Gross  and A.~A.~Tseytlin, ``Green-Schwarz string
  in AdS(5)  x  S(5): Semiclassical  partition  function,'' JHEP  {\bf
  0004}   (2000) 021   [arXiv:hep-th/0001204].    
  0001204;

\bibitem{Buscher:1987sk}
  T.~H.~Buscher,
  ``A Symmetry Of The String Background Field Equations,''
  Phys.\ Lett.\ B {\bf 194}, 59 (1987).

\bibitem{Bergshoeff:1995as}
  E.~Bergshoeff, C.~M.~Hull and T.~Ortin,
  ``Duality in the type II superstring effective action,''
  Nucl.\ Phys.\ B {\bf 451} (1995) 547
  [arXiv:hep-th/9504081].



\bibitem{Gibbons:2003rv}
  G.~W.~Gibbons and C.~E.~Patricot,
  ``Newton-Hooke space-times, Hpp-waves and the cosmological constant,''
  Class.\ Quant.\ Grav.\  {\bf 20}, 5225 (2003)
  [arXiv:hep-th/0308200].



\end{thebibliography}
\end{document}